\documentstyle[psfig,epsf,conf-X]{article}
\begin{document} 
\small
\heading{%
%Begin Heading
%
Cold clouds in cooling flows
\footnote{Poster presented for the International Workshop {\it Large scale 
structure in the X-Ray Universe}, Santorini Island (Greece), September 20-22$^{th}$, 1999}
% End Heading
}
\par\medskip\noindent
\author{%
%Begin Author names
Lukas Grenacher$^{1,2}$, Philippe Jetzer$^{1,2}$ and Denis Puy$^{1,2}$
%End Author names
}
\address{%
%First address
Paul Scherrer Institute, Laboratory for Astrophysics, 5232 Villigen and
}
\address{%
% Second Address
Institute of Theoretical Physics, University of Zurich, 8057 Zurich, Switzerland
}
%\address{%
% Third Address
% Here
%
%}

\begin{abstract}
In many clusters of galaxies there is evidence for cooling flows 
which deposit large quantities of cool gas in the central regions. A fraction 
of this gas might accumulate as dense cool clouds. The aim of this 
communication is to discuss the minimum temperature achievable by clouds 
in cooling flows of different clusters of galaxies.
\end{abstract}
\vskip2mm
\section{Molecular Cooling}
The ultimate fate of the gas which cools in cooling flows is still unknown. 
A possibility is 
that a fraction of the gas forms cold molecular clouds \cite{puy}. 
As a result of 
fragmentation we expect the formation of small clouds with higher density
and possible production of molecules in particular $H_2$ and traces of $HD$
and $CO$.\\
\newline
Taking into account radiative transfer effects we computed analytically 
in \cite{puy}
the cooling function $\Lambda(T)$ for the 
transition between the ground state and the first rotational level. This 
way we determined a lower limit of $\Lambda$.\\
Meanwhile we improved this calculation by including numerically all rotational 
transitions which are relevant at low temperatures (i.e. up to $J\sim 5$).\\
The following column densities are adopted for a typical small cloud (with
$n_{H_2}=10^6$cm$^{-3}$):
$N_{CO}=10^{14}\, {\rm cm}^{-2} \, \, {\rm and} \, \,  
N_{H_2}=2 \times 10^{18} \, {\rm cm}^{-2}$,
which corresponds to a $CO$ abundance:
$\eta_{CO}\sim 5 \times 10^{-5}$. For $HD$  instead we assume the 
primordial ratio $\eta_{HD} \sim 7 \times 10^{-5}$.\\
\newline
In Figure 1 we plotted the molecular cooling function $\Lambda(T)$ taking
into account $CO$, $HD$ and $H_2$ in 
the range 3-300 K. $CO$ is the main coolant in the range of temperatures 
3-80 K, $HD$ in the range 80-150 K and $H_2$ dominates above 150 K. 
\begin{figure}[h]
\centerline{\vbox{
\psfig{figure=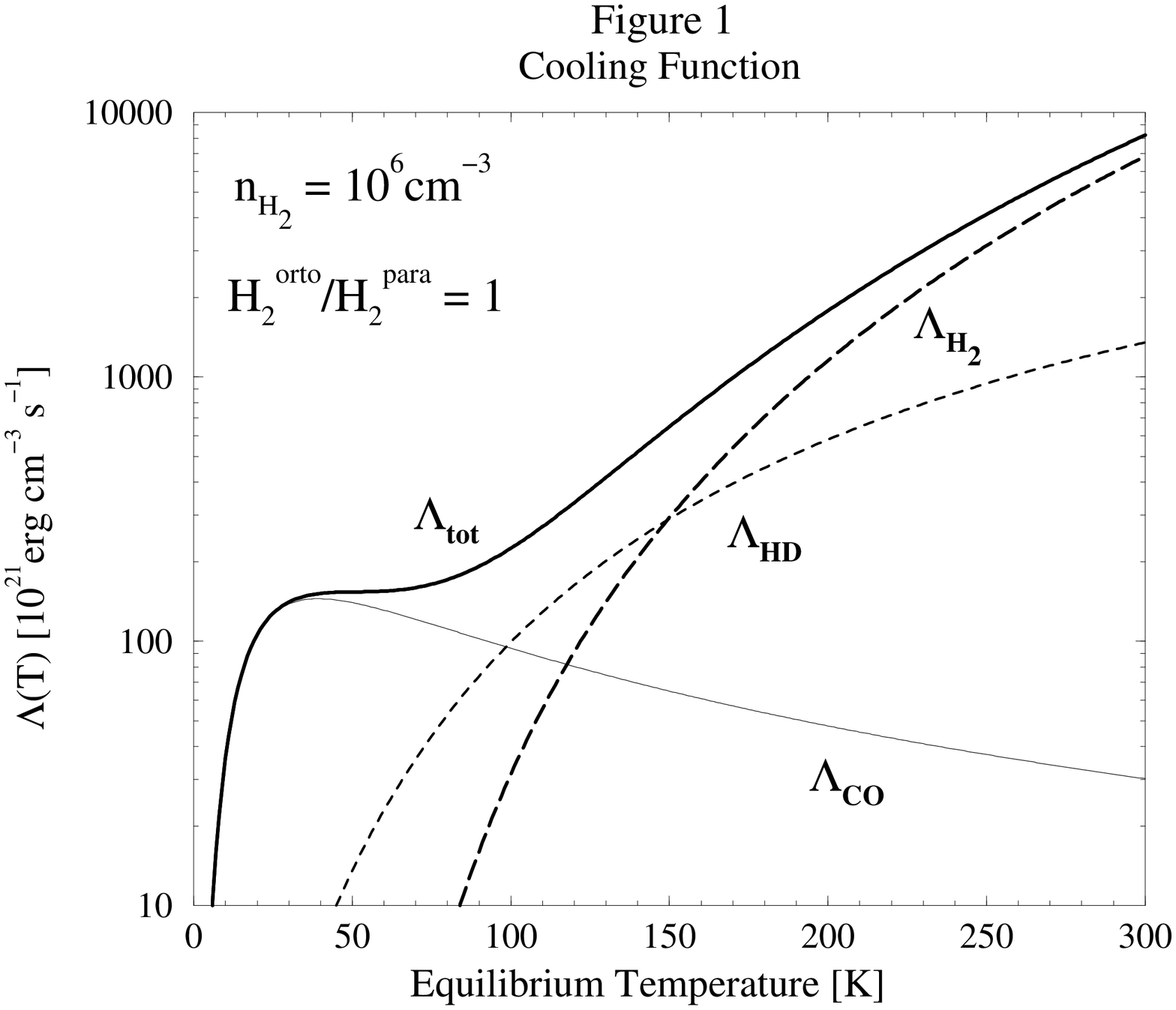,height=10.0cm,angle=-90}
}}
\end{figure}
\section{Thermal equilibrium}
The heating is given by the external X-ray flux as produced from the hot 
intracluster gas.
The thermal balance between heating and cooling 
leads to an equilibrium temperature for the clouds. We have calculated 
the minimum equilibrium 
temperature $T_{clump}$ of the clumps inside the cooling flow region. 
Table 1 shows the equilibrium temperature for different clusters.
For comparison we give the values we find using our analytical
approximation ($N$=1) and the ones by taking into account higher 
excited rotational levels ($N$=5).
One clearly sees that the inclusion of the higher excited levels into 
the calculations lowers the equilibrium temperature, particularly for
hot clusters such as for instance Abell 478.\\
\newline
We conclude that thermal equilibrium can be achieved at very low temperatures 
inside the cooling flow region mainly due to $CO$-cooling. Other molecules 
than $CO$, for example $CN$ or $H_2CO$, could also be important. Thus the 
study of the chemistry in cooling flows might lead to 
important insight.
\begin{center}
\begin{tabular}{l c c}
\multicolumn{3}{l}{{\bf Table 1.} Equilibrium temperature} \\
\hline
\multicolumn{1}{l}{Cluster}&\multicolumn{1}{c}{$T_{clump}~(N=1)$}&
\multicolumn{1}{c}{$T_{clump}~(N=5)$}\\
\multicolumn{1}{c}{ }&\multicolumn{1}{c}{(in K)}&
\multicolumn{1}{c}{(in K)}\\
\hline
PKS 0745-191 & 4  & 3 \\
Hydra A & 4 & 3 \\
Abell 478 & 75 & 10 \\
Centaurus & 25 & 4 \\
\hline
\end{tabular}
\end{center}
\acknowledgements{We would like to thank M. Plionis and I. 
Georgantopoulos for organizing such a pleasant conference. This work has been 
supported by the {\it Dr Tomalla Foundation} and by the Swiss National 
Science Foundation.}

\vfill
\end{document}